# Integrated Definition of Abstract and Concrete Syntax for Textual Languages

Holger Krahn, Bernhard Rumpe, and Steven Völkel

Institute for Software Systems Engineering
Technische Universität Braunschweig, Braunschweig, Germany
http://www.sse-tubs.de

**Abstract.** An understandable concrete syntax and a comprehensible abstract syntax are two central aspects of defining a modeling language. Both representations of a language significantly overlap in their structure and also information, but may also differ in parts of the information. To avoid discrepancies and problems while handling the language, concrete and abstract syntax need to be consistently defined. This will become an even bigger problem, when domain specific languages will become used to a larger extent. In this paper we present an extended grammar format that avoids redundancy between concrete and abstract syntax by allowing an integrated definition of both for textual modeling languages. For an amendment of the usability of the abstract syntax it furthermore integrates meta-modeling concepts like associations and inheritance into a well-understood grammar-based approach. This forms a sound foundation for an extensible grammar and therefore language definition.

## 1 Introduction

The definition of a language involves various kinds of activities. Usually a concrete and an abstract syntax is developed first, and then a semantics is designed to define the meaning of the language [9]. These activities are complemented by developing a type system, priorities for operators, naming systems etc. if appropriate. Especially the definition of concrete and abstract syntax show a significant redundancy, because domain concepts are reflected in both artifacts. This leads to duplications which are a constant source of problems in an iterative agile development of modeling languages. Despite general problems that occur when two documents are used, like inconsistency checking and supporting adequate ways to resolve them, domain concepts in the abstract syntax may be expressed on different ways in the concrete syntax. On the contrary the abstract syntax may contain elements that cannot be expressed in the concrete syntax of the language. These potential problems unnecessarily complicate an efficient development and evolution of languages, and therefore, an integrated development of both artifacts is highly desirable.

Meta-modeling is a popular method to define the abstract syntax of languages. It simplifies the language development by allowing the designers to directly map the classes of a domain analysis [2] to classes in the meta-model, associations and inheritance are directly part of the language definition. On the other side, grammar-based language definitions yield trees with single root-objects. Associations between leaves and an inheritance-based substitutability are not commonly existent in grammars.

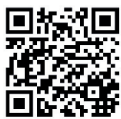

There are several approaches to derive a grammar and thus, a textual representation from a given metamodel. These approaches often lack flexibility in defining an arbitrary concrete syntax, it may even happen that the desired concrete syntax must be adapted in order to get an automatic mapping between abstract and concrete syntax (e.g., [11]). This stands in contrast to a basic design principle for DSLs [19] that already existing notations of the domain shall be used unaltered. Beyond that, one main argument for defining concrete syntax and abstract syntax separately is that more than a one concrete syntax can be used with a single abstract syntax. We argue, when dealing with DSLs this is a minor aspect because usually in a single domain no two notations are used that have the same expressiveness and therefore apply to same abstract syntax. However, we suggest the use of two similar abstract syntaxes and a (simple) model transformation for these rare cases.

The concrete syntax of a language is either texual or graphical. The graphical concrete syntax is often defined by the structure of the abstract syntax and a set of graphical representations for classes and associations in the abstract syntax (e.g. [7]). Especially for languages that do not have an adequate graphical representation, a textual syntax is used which is usually described by a context-free grammar. Parser-generators, e.g. Antlr [14] or SableCC [5] can be used to generate language recognition tools from this form of language definition. Since we aim at textual concrete syntaxes, we concentrate in this paper on the second approach.

The MontiCore framework [8] can be used for the agile development of textual languages, in particular domain-specific modeling languages (DSMLs). In order to reduce the abovementioned redundancy, one of the main design goals of the underlying grammar format was to provide a single language for specifying concrete as well as abstract syntax in a single definition. Associations, compositions, and inheritance as known from meta-modeling can directly be specified in this format. Such a language definition can be mapped to an object-oriented programming language where the each production is mapped to a class with strongly typed attributes. A parser is generated to create instances of the abstract syntax from a textual representation.

Despite these main design goals, we decided to develop the MontiCore grammar format in such a way, that the recognition power of the resulting parser is only limited by the underlying parser generator, namely Antlr [14], which is a predicated-LL(k) parser generator. Thus, it can not only be used for the development of domain specific modeling languages but for general-purpose languages like variants of Java or C++. The concrete syntax of the grammar format is similar to the input format of common parser-generators. Therefore, users that have already worked with such tools shall easily become familiar with it. The context-free grammars can be extended with meta-modeling concepts like associations and inheritance to define the abstract syntax of the modeling language.

The rest of the paper is structured as follows. Section 2 describes the syntax of the MontiCore grammar format and its semantics in form of the resulting concrete and abstract syntax of the defined modeling language. Section 3 describes an example that illustrates the clarity of the specification in the MontiCore grammar format. Section 4 relates our approach to others whereas Section 5 concludes the paper and outlines future work.

## 2  The MontiCore grammar format

The MontiCore grammar format specifies a context free grammar with productions that contain nonterminals (reference to other rules) and terminals. Terminals (also called identifiers) are specified by regular expressions. To simplify the development of DSLs, the identifer IDENT and STRING are predefined to parse names and strings. Identifiers are usually handled as strings, but more complex identifers are possible by giving a function defined in the programming language Java that maps a string to an arbitrary data type. Default functions exist for primitive data types like floats and integers. Examples are given in Figure 1. In line 2 the simple identifier IDENT is specified which is mapped to a String in the abstract syntax. The identifier NUMBER in line 5 is mapped to an integer in the abstract syntax whereat the default mapping is used. In line 8 the identifier CARDINALITY is mapped to an int. The transformation is specified in Java (line 10 and 11) and the unbounded cardinality is expressed as the value -1.

```
                       MontiCore-Grammar
 1 // Simple name
 2 ident IDENT ('a'..'z'|'A'..'Z')+ ;
 3
 4 // Numbers (using default transformation)
 5 ident NUMBER ('0'..'9')+ : int;
 6
 7 // Cardinality (STAR = -1)
 8 ident CARDINALITY ('0'..'9')+ | '*' :
 9   x -> int {
10     if (x.equals("*")) return -1;
11     else return Integer.parseInt(x);
12   };
```

**Fig. 1.** Definition of identifiers in MontiCore

The definition of a production in the grammar leads to a class with the same name in the abstract syntax. The nonterminals and identifiers on the right hand side of a rule can explicitly be given a name that maps to attribute names. For unnamed elements we derive default names from the name of the nonterminal.

The identifiers form the attributes of the class whereas the nonterminals lead to composition relationships between classes in the abstract syntax. The type of attributes in the abstract syntax is inferred automatically, the types of identifieres are handled as described before, attributes which form a composition relationship are typed with the class of the target rule. Thus, the attribute `name` of the rule `Client` in line 7 of Figure 2 results in a string attribute in the corresponding class of the abstract syntax, whereas `Address` in line 8 results in an attribute of the type `Address`. Additionally, the structure of a production is analyzed to determine the cardinality of the attributes and compositions. Doing so, attributes that occur more than once are realized as lists of the corresponding data type. This approach allows to specify constant separated lists without an extra construct in the grammar format. The term `a:X ("," a:X)*` can be used on the right hand side of a grammar rule and desribes a comma-separated list of the non-terminal `X`. It results in a composition named `a` with unbounded cardinality that contains all occurrences of `X`. Therefore, terminals and identifiers with the same name contribute to the same attribute or composition.

Figure 2 shows an excerpt of a MontiCore grammar. The class section shows a UML class diagram of the abstract syntax that is created from the productions. In the MontiCore framework this class diagram is mapped to a Java implementation where the production names are used as class names. All attributes are realized as private fields and public get- and set-methods. The composition relationships are realized in the same way as attributes and contribute to the constructor of the class. All classes support a variant of the Visitor pattern [6] to traverse the abstract syntax along the composition relationships.

In addition to the already explained nonterminals and identifiers, constant terminal symbols like keywords can be added to the concrete syntax of the language. These elements are not directly reflected in the abstract syntax if they are unnamed. Note that in contrast to many parser generators and languages like TCS [11] there is no specific need for distinguishing between keywords like "public" and special symbols like ",". To further simplify the development of a modeling language we generate the lexer automatically from the grammar. By this strategy the technical details like the distinction between parser and lexer (necessary for the parser generator) are effectively hidden from the language developer.

As explained above, keywords are not directly reflected in the abstract syntax as attributes, but may influence the creation of the AST by distinguishing productions with the same attributes from each other. The situation is different for reserved words that determine certain properties of domain concepts. An example is shown in Figure 3 where the reserved word *premiumclient* determines the value of an attribute of the domain concept client. The grammar format uses constants (inside brackets) to express this fact. Single value constants are translated to booleans whereas multi-value constants are mapped to enumerations.

The languages defined through the grammar in Figure 2 and the substituted grammar in Figure 3 are equal. But their abstract syntax is quite different. The concrete syntax poses the invariant that clients cannot have a discount whereas

```
                 ──────── MontiCore-Grammar ────────
 1  ShopSystem =
 2    name:IDENT
 3    (Client | PremiumClient)*
 4    (OrderCreditcard | OrderCash)*;
 5
 6  Client =
 7    "client" name:IDENT
 8    Address;
 9
10  PremiumClient =
11    "premiumclient"
12    name:IDENT discount:IDENT
13    Address;
14
15  OrderCreditcard =
16    "creditorder"
17    clientName:IDENT billingID:IDENT
18
19  OrderCash =
20    "cashorder"
21    clientName:IDENT amount:IDENT;
22
23  Address =
24    street:STRING town:STRING;
```

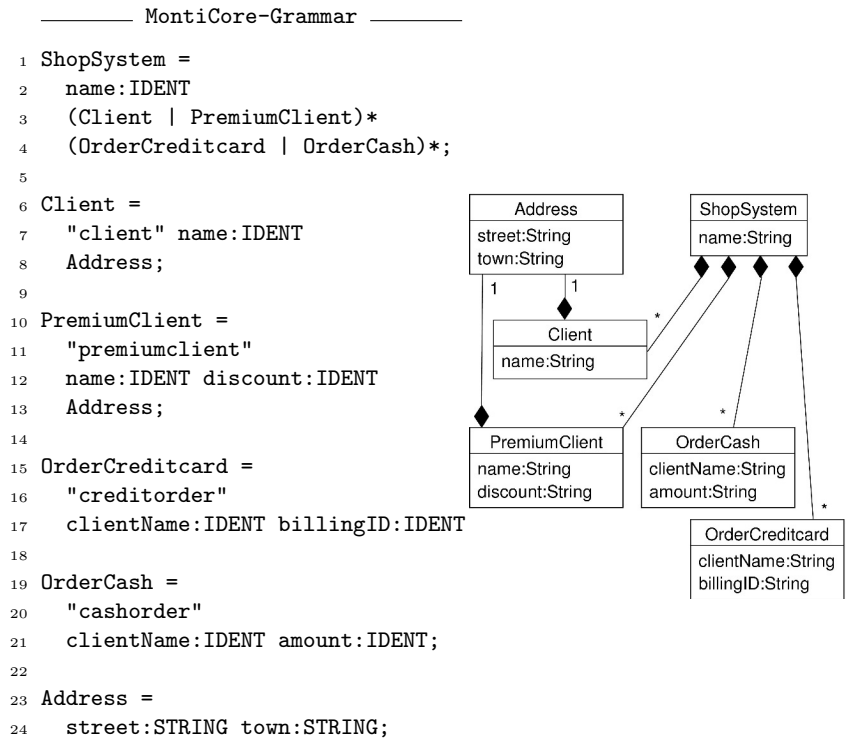

**Fig. 2.** Definition of productions in MontiCore

```
                 ──────── MontiCore-Grammar ────────
 1  Client =
 2    premium:["premiumclient"]
 3      name:IDENT discount:IDENT
 4    | "client" name:IDENT;
```

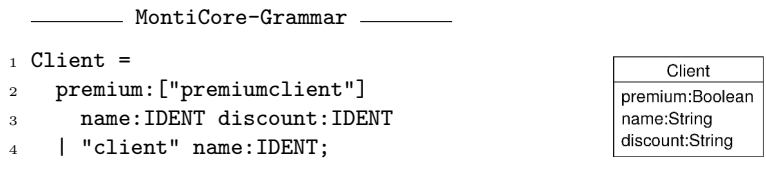

**Fig. 3.** Use of constants

premium clients do have one. This invariant is not visible in the abstract syntax from Figure 3. On the other hand, the abstract syntax resulting from Figure 2 doesn't reflect the similarities between Client and PremiumClient resp. Order-Cash and OrderCreditcard. This example motivates the use of more advanced features of the MontiCore grammar format to represent the invariants and similarities directly in the grammar. Despite very general constraint definitions (like OCL), inheritance allows us to deal with similarities and associations with connections between related nodes of the AST.

## 2.1 Interfaces and Inheritance between Nonterminals

The abstract syntax shown in Figure 2 raises the question how an interface `Order` that both classes `OrderCreditcard` and `OrderCash` implement can be added to the abstract syntax and how it can be expressed that premium clients are special clients. For this purpose we decided to extend the grammar format by allowing to express an inheritance relationship and to define interfaces which can be implemented by nontermianls.

The inheritance relationship between two productions is expressed by including the rule name of the super-production after the production name of the sub-production using the keyword `extends` (Figure 4, left, line 16). This inheritance of rules is directly reflected in the abstract syntax as an object-oriented inheritance. In the parser an additional alternative is added to the super-production. This concept is motivated by the definition of object-oriented inheritance where each occurrence of a superclass can be substituted by a subclass. The EBNF section in Figure 4, right, shows a representation with equivalent concrete syntax to explain the mapping of the MontiCore grammar format to the input format of a parser generator.

The grammar on the right hand side in Figure 4 defines the same concrete syntax as the one on the left, but has additional `Order` and `Client` rules. However, we have decided to use an OO style of inheritance instead of the traditional grammar style to get more flexibility in extending languages. In the left grammar, `Client` need not be changed when extending the language with `PremiumClients`. This is a significant benefit that we will further explore when defining operators on the language.

Due to experience in designing languages with this grammar format, we decided to decouple the concrete syntax of the both productions (sub- and super-production). This allows the language designer to decide freely on the concrete syntax and minimizes non-determinisms in the grammar.

This form of inheritance also allows the definition of superinterfaces using the keyword `implements` (Figure 4, left, line 5 and 9). Interfaces can be used as normal nonterminals on the right hand side of any production. By default a interface does not contain any attributes. We decided against an automatic strategy where all common attributes of known subclasses are taken, as the interface may be a good place for future extensions of the defined language which only use a subset of all available attributes.

Additional attributes may be added to interfaces and classes by using the `ast` section in a grammar (Figure 4, left, line 25). This block uses the same syntax as inside the production but only produces attributes in the abstract syntax and does not interfere with the concrete syntax. The attributes of interfaces are realized as get- and set-methods in the Java implementation. Figure 4 illustrates the inheritance capabilities of the grammar format extending the example. The EBNF section shows the equivalent EBNF syntax used for parsing and the resulting abstract syntax can be found in Figure 5.

```
     ──────── MontiCore-Grammar ────────              ──────── EBNF ────────
 1 ShopSystem =                                 1 ShopSystem ::=
 2   name:IDENT                                 2   IDENT Address
 3   Client* Order*;                            3   Client* Order*
 4                                              4
 5 OrderCreditcard implements Order =           5 OrderCreditcard ::=
 6   "creditorder"                              6   "creditorder"
 7   clientName:IDENT billingID:IDENT;          7   IDENT IDENT
 8                                              8
 9 OrderCash implements Order =                 9 OrderCash ::=
10   "cashorder"                               10   "cashorder"
11   clientName:IDENT amount:IDENT;            11   IDENT IDENT
12                                             12
13 Client =                                    13 Client ::=
14   "client" name:IDENT Address;              14   "client" IDENT Address
15                                             15   | PremiumClient ;
16 PremiumClient extends Client =              16
17   "premiumclient"                           17 PremiumClient ::=
18   name:IDENT discount:IDENT;                18   "premiumclient"
19                                             19   IDENT IDENT
20 Address =                                   20
21   street:STRING town:STRING;                21 Address ::=
22                                             22   STRING STRING
23 interface Order;                            23
24                                             24 Order ::=
25 ast Order =                                 25   OrderCredit | OrderCash
26   clientName:IDENT;                         26
     ─────────────────────────────              ─────────────────────────────
```

**Fig. 4.** Inheritance and use of interfaces

### 2.2 Associations between Nonterminals

The attributes `name` in `Client` and `clientName` in `Order` (see Figure 5) are obviously semantically connected. The invariant that an order may only use client names that exists cannot be expressed in a context-free grammar format. When designing a meta-model this relation is usually expressed by an association where an order references a client as the ordering person. Therefore, the extended context-free grammar allows to add associations and mimic typical meta-modeling techniques in grammars. The result of this extension is an arbitrary graph with an embedded spanning tree that results from the original grammar.

The block `association` allows to specify non-compositional associations between rules which enables the navigation between objects in the abstract syntax. This concept allows to specify a uni-directed navigation from one object to a specified number of other objects. In addition, an opposite association can be specified that reverses the first association. An example for an association can be found in Figure 6 (line 17-21) where the association OrderingClient connects one Order object with a single Client. The reverse association is named Order (the name is automatically derived from the target) which connects one Client object with an unbounded number of Order objects. This form is very similar to the associations in EMF [1].

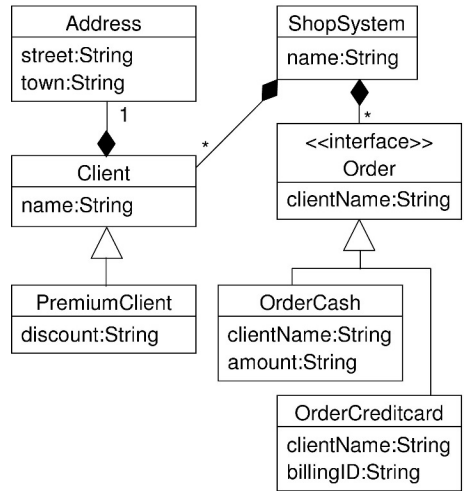

**Fig. 5.** Abstract syntax of the language defined in Figure 4

The main challenging question for associations in a unified format for concrete and abstract syntax is not the specification but to automatically establish the links between associated objects from a parseable textual input. Grammar-based systems usually parse the linear character stream and represent it in a tree-based structure that has the same structure as the grammar. Then symbol tables are used to navigate between nodes in the AST that are not directly connected. The desired target of navigation is determined by identifiers in source and target nodes and a name resolution algorithm.

Due to the simple nature of many languages that lack namespaces, simple resolution mechanisms like file-wide unique identifiers can often be used for an establishment of associations. This of course does not always work. E.g., languages like Java and many UML-sublanguages do have a more sophisticated namespace concept.

Therefore, we decided to use a twofold strategy: First, we generate interfaces that contain methods induced by the association to navigate between the AST-objects. The resulting classes of the abstract syntax allow the access of associations in the same way as attributes and compositions are accessed. Second, we generate implementations for simple resolving problems like file-wide flat simple or hierarchical namespaces. As an alternative, the DSL developer can program his own resolving algorithms in the second step if needed.

Figure 6 extends the example from Figure 5 by adding an association specification. The association `orderingClient` connects each `Order` to a single Client (as specified by `1`). `Order` is the inverse association from Client to Order with unbound cardinality (as specified by `*`). In addition to the shown cardinalities, ranges like `3..4` are possible values.

```
                ──── MontiCore-Grammar ────
1  OrderCreditcard implements Order =
2    "creditorder"
3    iD:IDENT amount:IDENT;
4
5  OrderCash implements Order =
6    "cashorder"
7    iD:IDENT amount:IDENT;
8
9  Client =
10   "client" name:IDENT
11   Address;
12
13 PremiumClient extends Client =
14   "premiumclient"
15   name:IDENT discount:IDENT;
16
17 association {
18   Order.orderingClient
19   * <-> 1
20   Client
21 }
```

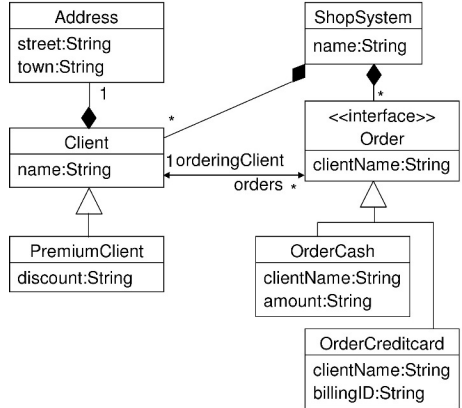

**Fig. 6.** Specification of associations

Figure 7 sketches a Java implementation of the class diagram from Figure 6 with the most important methods. A Binding-Interface is generated for each interface or class that is involved in an association as either source or target. This interface contains the relevant methods for the navigation between different nodes. In addition a Resolver is generated for each class or interface which allows the resolving of a Binding-object from an AST-object.

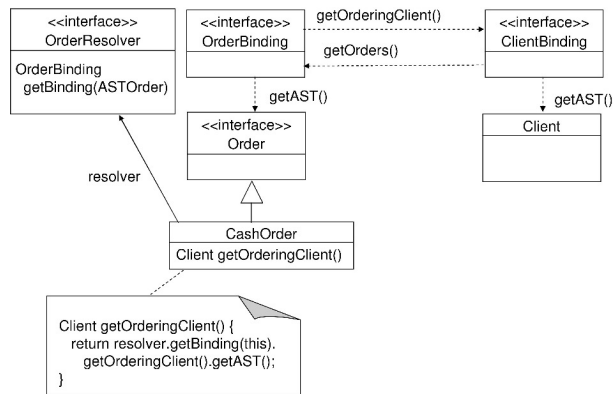

**Fig. 7.** Java implementation of an association

Note that these interfaces are generated to simplify the use of associations for a DSL. When simple resolving algorithms are appropriate, MontiCore can generate both Binding-implementations and a single Resolver-implementation that resolves all objects automatically. The complexity of multiple classes with different responsibilities is hidden from the user of the abstract syntax, e.g. a programmer of a code generator for the developed language. He simply uses the get- and set-methods like the shown getOrderingClient() method that returns the client object which is referred from this order.

## 3  A demonstrating Example

To demonstrate the usability of our approach to specify a modeling language we use a simplified version of finite hierarchical automata as shown in Figure 8.

─────────────── Automaton ───────────────
```
1  automaton PingPong {
2    state <<initial>> NoGame;
3    NoGame - startGame > InPlay;
4    InPlay - ["doStopGame()"] stopGame > NoGame;
5    state InPlay {
6      state <<initial>> Ping;
7      state Pong;
8    }
9    Ping - returnBall > Pong;
10   Pong - returnBall > Ping;
11 }
```
─────────────────────────────────────────

**Fig. 8.** Example for finite hierarchical automata

An automaton has a name and consists of several states and transitions. States in turn may be initial or final and may have substates. Transitions have a source and a target state, an event models the condition for the transition. In order to demonstrate a possible field of application for grammar rule inheritance, a transition may have an action which will be executed when a transition was performed. Figure 9 shows a first version of the MontiCore grammar.

In our example transitions refer to states as source and target which will be identified by their name. Thus, the generated class Transition contains string attributes from and to containing the names of these states. This is ineffective because a direct navigation from transitions to their source or target is not possible. Furthermore, there is no information in states about their ingoing and outgoing transitions. The abovementioned concept association can be used to solve this. An appropriate association extending Figure 9 is defined in Figure 10.

This definition leads to attributes which refer to states and transitions directly. They have to be filled by an appropriate resolve mechanism depending

```
                          MontiCore-Grammar
1 package mc.languages.automaton;
2
3 grammar Automaton {
4
5   Automaton = !"automaton" name:IDENT "{"
6     ( State | Transition )*
7   "}";
8
9   State =
10    !"state" name:IDENT
11    ( "<<" initial:[!"initial"] ">>" | "<<" final:[!"final"] ">>" )*
12    ( "{" State* "}" | ";" ) ;
13
14  Transition = from:IDENT "-" event:IDENT ">" to:IDENT ";";
15
16  TransitionWithAction extends Transition =
17    from:IDENT "[" action:STRING "]"  "-" event:IDENT ">" to:IDENT ";";
18 }
```

**Fig. 9.** Basic definition of finite hierarchical automata in MontiCore

```
                          MontiCore-Grammar
1   association {
2     Transition.fromState * <-> 1 State.outgoingTransitions * ;
3     Transition.toState * <-> 1 State.incomingTransitions * ;
4   }
```

**Fig. 10.** Definition of non-compositional associations between states and transitions

on the underlying naming system. As described in Section 2.2, MontiCore supports different kinds of resolve mechanisms, in this example we use the simplest version, namely file-wide unique identifiers which is defined by the concept simplereference. Therefore, states must have a unique name within the automaton and transitions use that name in order to reference these states. Beyond that, we have to specify that our concept for simple references should be used in order to resolve the associations defined in Figure 10. Therefore, the code of Figure 11 has to be added to our automaton grammar (Figure 9).

```
                          MontiCore-Grammar
1   concept simplereference {
2     fromState: Transition.from -> State.name;
3     toState: Transition.to -> State.name;
4   }
```

**Fig. 11.** Using concept simplereference to resolve source and target of transitions

Given both, the definition of simple references and the concept association, the MontiCore framework ensures the following constraints.

1. Each transition refers exactly one state as source and target, respectively.
2. Each referenced state has a unique name within all referenced states.
3. The method getIncomingTransitions() of a state returns all transitions which refer to that state as target. Therefore, `incomingTransitions` is the opposite association of ToState.
4. The method getOutgoingTransitions() of a state returns all transitions which refer to that state as source. Therefore, `outgoingTransitions` is the opposite association of FromState.

Another useful feature we want to present in this example is the concept classgen. It can be used in order to add attributes as well as methods into the abstract syntax. Therefore, it has to be defined which class should be extended and what should be added to that class. Again, the example shown in Figure 12 can be added to the basic grammar (Figure 9). It adds the boolean method isDirectlyReachable to the abstract syntax class State which calculates if one state is directly reachable from this state.

―――――――――――――――――― MontiCore-Grammar ――――――――――――――――――
```
1 ast State =
2   method public boolean isDirectlyReachable(State target) {
3     for (Transition t: getOutgoingTransitions()){
4       if (t.getToState().equals(target)){
5         return true;
6       }
7     }
8     return false;
9   };
```

**Fig. 12.** Using an ast block to add methods to states

Summarizing, we have developed a grammar for finite hierarchical automata in a few lines of code. Non-compositional bi-directional associations are supported and filled automatically by a simple naming system which ensures correct cardinalities. An example of an additional method defined in Java enhances usability as well as it prevents editing the generated code.

## 4 Related work

We are currently not aware of a language that allows specifying both a textual concrete syntax and an abstract syntax with (non-compositional) associations in a coherent and concise format. Grammar-based approaches usually lack a strongly typed internal representation (for exceptions see below) and the existing model-based approaches use two forms of description, a meta-model for the abstract syntax and a specific notation for the concrete syntax.

In [15] a phylum/operator-notation is used to describe the abstract syntax of a language. The notation of alternate phylums achieves similar results as the object-oriented inheritance we use, although our tied coupling of the abstract syntax to a programming language allows the direct use of the inheritance to simplify the implementation of algorithms working on the abstract syntax.

SableCC [5] is a parser-generator that generates strictly-typed abstract syntax trees and tree-walkers. The grammar format contains actions to influence the automatic derivation of the AST. In contrast to MontiCore, SableCC does not aim to include associations in its AST.

In [18] an algorithm is presented that derives an (strongly typed) abstract syntax from a WBNF grammar (an BNF variant). The main difference in the derivation to our approach is the use of an explicit notation for lists that are separated by constants and that nonterminals with same name do not contribute to the same attribute in the abstract syntax.

The Grammar Deployment Kit (GDK) [12] consists of several components to support the development of grammars and language processing tools. The internal grammar format can be transformed into inputs of different parser generators, such as btyacc [3], Antlr [14] or SDF [10]. Furthermore, it provides possibilities for grammar adaption, like renaming of rules or adding alternatives. In opposition to our approach it does not support extended concepts like inheritance or associations.

In [4] and [13] the Textual Concrete Syntax Specification Language (TCSSL) is described that allows the description of a textual concrete syntax for a given abstract syntax in form of a meta-model. TCSSL describes a bidirectional mapping between models and their textual representation. The authors describe tool support to transform a textual representation to a model and back again.

In [11] a DSL named TCS (Textual Concrete Syntax) is described that specifies the textual concrete syntax for an abstract syntax given as a meta-model. The described tool support is similar to the one we used for the MontiCore framework and the name resolution mechanisms are the same that we generate automatically from the grammar format. In contrast to our approach, two descriptions for abstract and concrete syntax are needed.

## 5 Conclusion

This work presents a new approach where an extended grammar format is used to specify both, abstract and concrete syntax of a modeling language. By using a single format it avoids general problems that occur when abstract and concrete syntax are described by two different languages like inconsistency checking and resolving by construction.

As special concepts, we added the possibility to define associations between AST nodes based on name references and we allow inheritance of grammar rules that does not affect the super-rule at all. This paves the way for extensible languages.

We also implemented a prototypical framework called MontiCore that is based on an established parser-generator. It is able to parse textual syntax and generates the model representation as both Java and EMF classes. The prototype is able to parse multiple language definitions like UML/P [17, 16] and a complete Java 5 grammar. In addition the system is bootstrapped and currently about 75% of the code is generated from several DSLs.

In future we especially want to explore which resolution mechanisms can be used to create links between objects (that conform to the specified associations). The mechanisms for resolving imports in models and inheritance seem to be promising candidates for generalization.

We mainly treated the transformation from concrete syntax to abstract syntax representation in this paper. The opposite transformation, where a model is transformed to a concrete (textual) representation could be realized in different ways: Hand coded java code, template engines etc. In the future we like to explore which additional information has to be included in the grammar format to allow the automatic generation of concrete syntax representations.

The current implementation of the parser generation in the MontiCore framework is based on Antlr version 2.x. The new version 3 simplifies the creation of grammars by automatically calculating the necessary syntactic predicates for alternatives where the linearized lookahead algorithm predicts false results. This will reduce the number of required syntactic predicates and simplify the development of readable grammars in MontiCore.

**Acknowledgement:** The work presented in this paper is undertaken as a part of the MODELPLEX project. MODELPLEX is a project co-funded by the European Commission under the "Information Society Technologies" Sixth Framework Programme (2002-2006). Information included in this document reflects only the authors' views. The European Community is not liable for any use that may be made of the information contained herein.